\documentclass[aps,prl,twocolumn,showpacs,superscriptaddress,groupedaddress]{revtex4}  

\usepackage{rotating}

\usepackage{natbib}
\usepackage{amsmath}
\usepackage{graphicx}
\usepackage{dcolumn}
\usepackage{bm}
\newcommand{\kp}{k_\perp}
\newcommand{\beq}{\begin{equation}}
\newcommand{\eeq}{\end{equation}}
\newcommand{\bkp}{b_{\kp}}
\newcommand{\ekp}{E_{\kp}}
\newcommand{\onl}{\omega_{nl}}
\newcommand{\tok}{\tilde{\omega}_{\kp}}
\newcommand{\epskp}{\epsilon_{\kp}}
\newcommand{\gkp}{\gamma_{\kp}}
\begin{document}

\title{Inner-Heliosphere Signatures of Ion-Scale Dissipation and Nonlinear Interaction}

\author{Trevor A. Bowen}\email{tbowen@berkeley.edu}

\affiliation{Space Sciences Laboratory, University of California, Berkeley, CA 94720-7450, USA}
\author{Alfred Mallet}
\affiliation{Space Sciences Laboratory, University of California, Berkeley, CA 94720-7450, USA}
\author{Stuart D. Bale}
\affiliation{Space Sciences Laboratory, University of California, Berkeley, CA 94720-7450, USA}
\affiliation{Physics Department, University of California, Berkeley, CA 94720-7300, USA}
\affiliation{The Blackett Laboratory, Imperial College London, London, SW7 2AZ, UK}
\affiliation{School of Physics and Astronomy, Queen Mary University of London, London E1 4NS, UK}
\author{J. W. Bonnell}
\affiliation{Space Sciences Laboratory, University of California, Berkeley, CA 94720-7450, USA}
\author{Anthony W. Case}
\affiliation{Smithsonian Astrophysical Observatory, Cambridge, MA 02138 USA}
\author{Benjamin D. G. Chandran}
\affiliation{Department of Physics \& Astronomy, University of New Hampshire, Durham, NH 03824, USA}
\affiliation{Space Science Center, University of New Hampshire, Durham, NH 03824, USA}
\author{Alexandros Chasapis}
\affiliation{Laboratory for Atmospheric and Space Physics, University of Colorado, Boulder, CO 80303, USA}
\author{Christopher H. K. Chen}
\affiliation{School of Physics and Astronomy, Queen Mary University of London, London E1 4NS, UK}
\author{Die Duan}
\affiliation{Space Sciences Laboratory, University of California, Berkeley, CA 94720-7450, USA}
\affiliation{School of Earth and Space Sciences, Peking University, Beijing, 100871, China}

\author{Thierry {Dudok de Wit}}
\affiliation{LPC2E, CNRS and University of Orl\'eans, Orl\'eans, France}
\author{Keith Goetz}
\affiliation{School of Physics and Astronomy, University of Minnesota, Minneapolis, MN 55455, USA}
\author{Jasper S. Halekas}
\affiliation{Department of Physics and Astronomy, 
University of Iowa, 
Iowa City, IA 52242, USA}

\author{Peter R. Harvey}
\affiliation{Space Sciences Laboratory, University of California, Berkeley, CA 94720-7450, USA}
\author{J. C. Kasper}
\affiliation{Climate and Space Sciences and Engineering, University of Michigan, Ann Arbor, MI 48109, USA}
\affiliation{Smithsonian Astrophysical Observatory, Cambridge, MA 02138 USA}
\author{Kelly E. Korreck}
\affiliation{Smithsonian Astrophysical Observatory, Cambridge, MA 02138 USA}

\author{Davin Larson}
\affiliation{Space Sciences Laboratory, University of California, Berkeley, CA 94720-7450, USA}
\author{Roberto Livi}
\affiliation{Space Sciences Laboratory, University of California, Berkeley, CA 94720-7450, USA}
\author{Robert J. MacDowall}
\affiliation{Solar System Exploration Division, NASA/Goddard Space Flight Center, Greenbelt, MD, 20771}
\author{David M. Malaspina}
\affiliation{Laboratory for Atmospheric and Space Physics, University of Colorado, Boulder, CO 80303, USA}
\author{Marc Pulupa}
\affiliation{Space Sciences Laboratory, University of California, Berkeley, CA 94720-7450, USA}
\author{Michael Stevens}
\affiliation{Smithsonian Astrophysical Observatory, Cambridge, MA 02138 USA}
\author{Phyllis Whittlesey}
\affiliation{Space Sciences Laboratory, University of California, Berkeley, CA 94720-7450, USA}

\begin{abstract}
We perform a statistical study of the turbulent power spectrum at inertial and kinetic scales observed during the first perihelion encounter of Parker Solar Probe. We find that often there is an extremely steep scaling range of the power spectrum just above the ion-kinetic scales, similar to prior observations at 1 AU, with a power-law index of around $-4$. Based on our measurements, we demonstrate that either a significant ($>50\%$) fraction of the total turbulent energy flux is dissipated in this range of scales, or the characteristic nonlinear interaction time of the turbulence decreases dramatically from the expectation based solely on the dispersive nature of nonlinearly interacting kinetic Alfv\'en waves.
\end{abstract}

\maketitle
\paragraph{Introduction.---} In many astrophysical settings, the background plasma is both highly turbulent and nearly collisionless. The dissipation of this collisionless turbulence is important for heating the plasma \citep{Richardson1995,Quataert1998,Cranmer2000,CranmervanBallegooijen2003,Zhuravleva2014,Chen2019}, but the precise physical mechanisms involved are still a matter of debate \citep{Parashar2015}. As an example, the observed ion temperature profiles in the solar wind requires significant (perpendicular) ion heating, which is likely initiated at around the ion scales where particles can interact efficiently with electromagnetic waves \citep{Chandran2011}. Such heating should cause a transfer of energy from the waves to the particles around the ion scales, and thus would cause a steepening of the spectrum at these scales; this motivates our present study.

The solar wind provides a convenient example of collisionless plasma turbulence that can be studied using \emph{in situ} spacecraft observations. Taylor's hypothesis $\onl \ll \kp V_{\rm SW}$, where $V_{\rm SW}$ is the solar wind speed, is assumed to be well satisfied, so that the observed spacecraft-frame frequency spectra may be simply converted to wavenumber spectra. At large scales (much larger than characteristic ion-kinetic scales), the dominant turbulent fluctuations appear to be nonlinearly-interacting Alfv\'enic turbulence \citep{Chen2016}, with a power-law spectrum between $\kp^{-5/3}$ \citep{Matthaeus1982a} and $\kp^{-3/2}$ \citep{Podesta2007,Chen2019}, in rough agreement with various MHD turbulence theories \citep{GS95,Boldyrev2006,Chandran2015,MalletSchekochihin2017}. At scales much smaller than the ion gyroradius $\rho_i = v_{thi}/\Omega_i$ (with $v_{thi}=\sqrt{2T_{0i}/m_i}$ the ion thermal speed and $\Omega_i=ZeB_0/m_i$ the ion gyrofrequency), the spectrum steepens to about $\kp^{-2.8}$ \citep{Alexandrova2008,Sahraoui2009,Chen2010a}, as the non-dispersive Alfv\'enic turbulence transition to dispersive kinetic Alfv\'enic turbulence, as confirmed, for example, by measurements of the density fluctuation spectrum \citep{Chen2013}. This steepening occurs due to the change in the dispersion relation, and occurs even without any dissipation: a fluid approximation to the dynamics in this range of scales (ERMHD) leads to a prediction of an $\kp^{-7/3}$ spectrum \citep{Schekochihin2009}, while simulations of this fluid approximation obtain a $k_\perp^{-8/3}$ spectrum \citep{BoldyrevPerez2012}, which has been ascribed to intermittency. Our results will broadly confirm this picture of the large- and small- scale turbulent spectrum.

In addition to these power-law scalings at large and small scales, a ``transition range" around the ion scales with a spectrum significantly steeper than the $\kp^{-2.8}$ in the sub-ion range has often been observed in the solar wind \citep{Denskat1983,Leamon1998a,Smith2006,Sahraoui2010,Kiyani2015,Lion2016}. Because this anomalous steepening disappears at scales deeper in the sub-ion range, it is not possible to explain by means of the kinetic Alfv\'en wave (KAW) dispersion relation. The two main proposed explanations are, first, strong dissipation of the turbulence around the ion scales \citep{Chandran2010}, and, second, nonlinear effects which may increase the characteristic cascade rate of the turbulence: e.g., the onset of reconnection \citep{Mallet2017b,Loureiro2017,Vech2018}, the increasing importance of nonlinear interactions between co-propagating waves \citep{Voitenko2016}, or the influence of coherent structures \citep{Alexandrova2008b,Lion2016,Perrone2016}. In this Letter, we assess these two possibilities using a simple model which is agnostic as to the exact physical mechanisms responsible for dissipation and increased cascade rates.
\paragraph{Cascade model.---} We use a Batchelor cascade model \citep{Batchelor1953,Howes2008}. The turbulent energy flux $\epskp$ through wavenumber $\kp$ is related to the spectrum $\ekp= \bkp^2/\kp$, $\bkp$ being the turbulent amplitude at $\kp$, via
\beq
\epskp = \onl\kp \ekp,\label{eq:epsE}
\eeq
where
\beq
\onl = \kp \bkp \tok = \kp^{3/2}\ekp^{1/2}\tok\label{eq:onl}
\eeq
is the characteristic nonlinear frequency \footnote{We have absorbed all \emph{a priori} unknown constants of order unity into the definitions of $\onl$ and $\tok$.} (inverse cascade time) at $\kp$, and $\tok$ parametrizes both dispersive (e.g., from the dispersive KAW \citep{Schekochihin2009}) and/or nonlinear effects (e.g. caused by dynamic alignment \citep{Boldyrev2006},  intermittency \citep{BoldyrevPerez2012}, or reconnection \citep{Mallet2017b}). In statistical steady state far from the injection scales (cf. \citep{Howes2008,Howes2011}) one obtains a simple equation with solution
\beq
1-\hat{Q}_{k_1,k_2}=\frac{\epsilon_{k_2}}{\epsilon_{k_1}} = \exp{\left\{-\int_{k_1}^{k_2} \frac{\gkp}{\onl}\frac{d\kp}{\kp}\right\}},
\eeq
where $\gkp$ is the energy dissipation rate at $\kp$ (we do not specify a physical mechanism), and $\hat{Q}_{k_1,k_2}$ is the fractional heating rate over the range $[k_1,k_2]$. If $\gkp=0 \,\forall\, \kp \in [k_1,k_2]$, then $\epskp$ is conserved between $k_1$ and $k_2$.  Using Eqs.~\ref{eq:epsE}-\ref{eq:onl},
\beq
\frac{E_{k_2}}{E_{k_1}} = \left(\frac{k_2}{k_1}\right)^{-5/3}\left(\frac{\epsilon_{k_2}}{\epsilon_{k_1}}\right)^{2/3}\left(\frac{\tilde{\omega}_{k_1}}{\tilde{\omega}_{k_2}}\right)^{2/3}.\label{eq:Eeps}
\eeq
Deviations from a $k^{-5/3}$ spectrum (first bracket) must be caused by either dissipation (second bracket) or dispersive/nonlinear effects (third bracket) \footnote{Two examples of the latter are the following: first, deep in the subion range the linear dispersion of KAW should impose $\tok \sim \kp\rho_i$, which steepens the spectrum to the previously mentioned $\kp^{-7/3}$. Second, incorporating dynamic alignment at large scales \citep{Boldyrev2006} provides $\tok \propto \kp^{-1/4}$, which gives the expected $\kp^{-3/2}$ spectrum.}. In our analysis, we use Eq.~\ref{eq:Eeps} to relate \emph{in situ} measurements of the spectrum to estimates of the heating rate and/or anomalous $\tok$ scalings in the transition range.

\paragraph{Data and fitting.---}
The FIELDS \citep{Bale2016} and Solar Wind Electron Alpha and Proton (SWEAP, \cite{Kasper2016}) instrument suites on the Parker Solar Probe (PSP) mission \citep{Fox2016} provide {\em{in situ}} measurements of the inner-heliosphere plasma environment, enabling detailed studies of turbulence \citep{Chen2019,Martinovic2019,McManus2019}. Preliminary observations using MAG data show a steep $f^{-3}$ to $f^{-4}$ spectrum of the magnetic field fluctuations close the the ion scales \citep{Bale2019,Duan2019,Vech2019}, but the noise floor of the MAG at higher frequencies has so far precluded a detailed study.

In this Letter, we use a data-set which merges the PSP/FIELDS MAG and SCM measurements \cite{Bowen2020}, operating at 293 samples per second, enabling simultaneous measurements of the full range of ineritial, transition, and kinetic turbulent scales. Our data is from the first PSP perihelion encounter, when the spacecraft was magnetically connected to a small equatorial coronal hole generating slow, but highly Alfvenic solar wind \cite{Bale2019}, from 2018-11-04/09:28:19 to 2018-11-07/09:28:19. We separate the encounter into intervals of $2^{16}$ samples ($\sim$223.69 s). A $50\%$ overlap between intervals is used improve statistics. Average plasma $n_0$, $T_{0i}$, and $T_{0e}$ are computed for each interval using the SWEAP data \cite{Kasper2016,Case2020,Halekas2020}. The spectral density is computed by averaging eight non-overlapping sub-intervals of vector magnetic field measurements. 
 Intervals were rejected if no finite SWEAP measurements exist, or if the SCM was in a low gain mode ($\sim$ 1 hr each day). Intervals with ion scale waves, observable in 30-50\% of radial field intervals, which strongly affect measurements of ion-scale turbulence, are excluded \cite{Bale2019,Bowen2019a}. In total, 227 intervals were kept. We consider frequencies up to 100 Hz, corresponding to $\kp\rho_i\approx 10$, avoiding the SCM noise floor which is occasionally reached at higher frequencies.

KAW have intrinsic density fluctuations  \citep{Howes2006,Schekochihin2009}, which at $\kp\rho_i\gtrsim 1$ provide a non-negligible contribution to the total free energy. We estimate this contribution by determining $\delta n_e$ from the pressure balance $\delta B_\parallel/B_0 = -(\beta_i/2) (1+ZT_{0e}/T_{0i}) \delta n_e/n_{0e}$, appropriate for KAW, and estimate the total free energy \footnote{No significant impact on the results is obtained when the trace magnetic-field spectrum is used without this correction; this is likely because all the relevant quantities scale in the same way with wavenumber.} as
\beq
E_{\rm tot} = \frac{|\delta\bf{B}|^2}{2\mu_0}+\frac{n_{0e}T_{0e}}{2}\left(\frac{\delta n_e}{n_0}\right)^2.
\eeq

\begin{figure}
    \centering
    \includegraphics[width=3.375in]{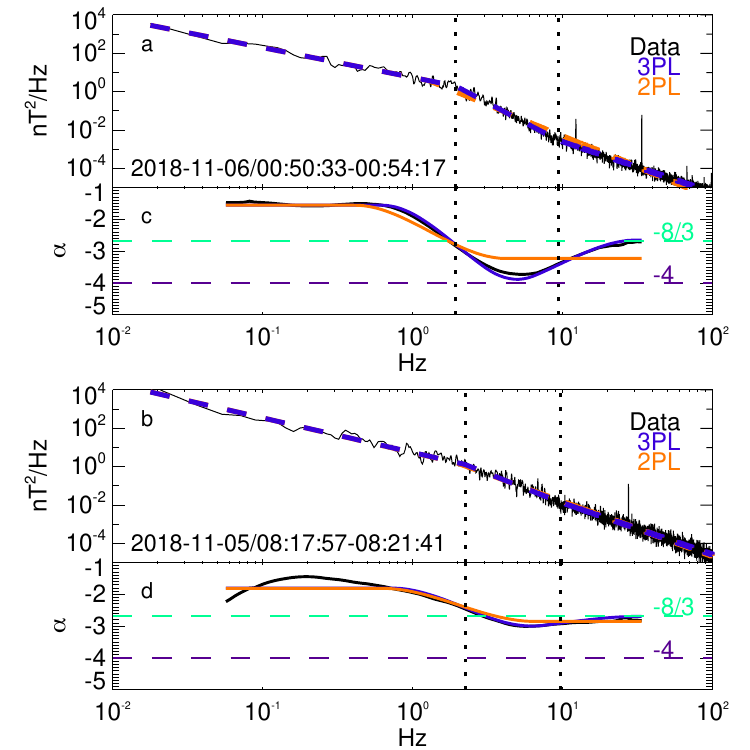}
   \caption{(a,b) Examples of PSP/FIELDS magnetic field spectra with 3PL (blue) and 2PL fits (orange). Black dashed lines show measured 3PL spectral breaks. (c,d) spectral indices computed from data (black), 3PL (blue) and 2PL fits (orange). Dashed lines are shown corresponding to spectral indices of -8/3 (teal) and -4 (purple).  Top interval has statistically significant spectral steepening, while the bottom interval is well modeled by the 2PL fit.}
    \label{fig:1}
\end{figure}
Figure \ref{fig:1}(a) shows an example interval with transition-range steepening to an approximate $f^{-4}$ spectrum at ion scales, similar to Cluster observations at 1 AU \cite{Sahraoui2010, Kiyani2015}. At the highest frequencies, the measured spectral index is consistent with the modified kinetic Alfv\'{e}n wave scaling of $E\propto f^{-8/3}$. Anomalous transition-range steepening is not present in a second example interval, presented in Figure \ref{fig:1}(b), but an approximate $f^{-8/3}$ scaling at the higher frequencies is evident.

Spectra are fit with two-power-law (2PL) and three-power-law (3PL) functions using non-linear least square optimization of $\chi^2$ residuals. Figure \ref{fig:1} shows the local moving window spectral index computed over a decade of frequencies for the data and each model (Figure \ref{fig:1}(c,d). 
The 3PL fit allows for determination of spectral indices of the inertial ($\alpha_I$), transition, ($\alpha_T$), and kinetic ($\alpha_K$) ranges and the break points ($f^{*}_{IT}$ and $f^{*}_{TK}$). The inclusion of a third spectral range introduces two additional degrees of freedom (the transition break, $f^*_{IT}$, and index, $\alpha_T$), which inherently improve the square residuals \footnote{the set of all 2PL fits is within the span of 3PL fits}. The significance of the improvement in $\chi_{3PL}^2$ (over $\chi_{2PL}^2$) when removing degrees of freedom (DOF), e.g. through including more fit parameters, is determined by the probability $P(F)$ of drawing $F$ from the appropriate $f$-distribution \citep{Press}:  
$$ F=\frac{\chi^2_{2PL}-\chi^2_{3PL}}{DOF_{2PL}-DOF_{3PL}}/\frac{\chi^2_{3PL}}{DOF_{3PL}}.$$

\begin{figure}
   \centering
    \includegraphics[width=3.375in]{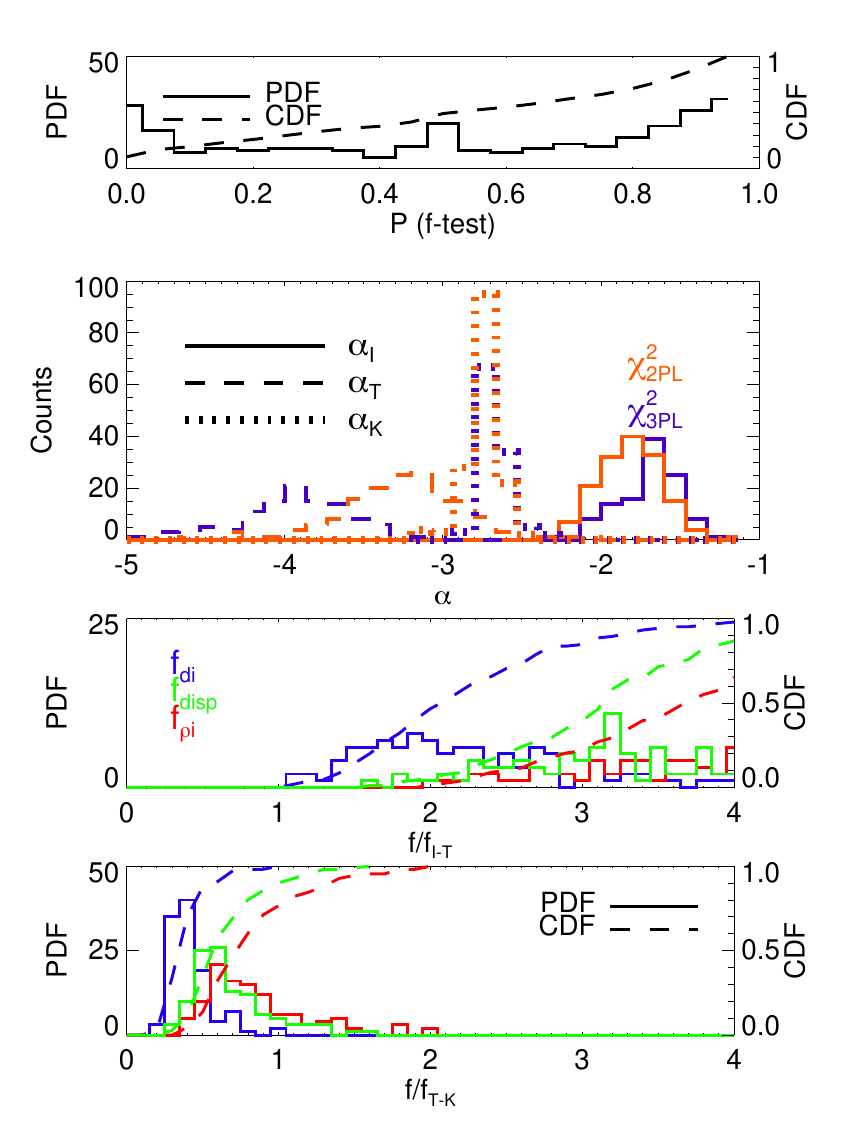}
    \caption{(a) Histogram of $f$-test significance values $P(F)$; the cumulative distribution is shown as a dashed line. (b) Distribution of fitted spectral indices for the $\chi^2_{3PL}$ (blue) and $\chi^2_{2PL}$ (orange) populations, in the inertial $\alpha_I$ (solid), transition $\alpha_T$ (dashed), and kinetic $\alpha_K$ (dotted) ranges. (c,d) Measured break  frequencies ($f_{IT}$,$f_{TK}$) compared to physical plasma scales: $f_\rho$ (red), $f_{di}$ (blue), and $f_{disp}$ (green).}
    \label{fig:2}
\end{figure}

Figure \ref{fig:2}(a) shows the distribution of measured $f$-test significance values, $P(F)$.  The continuum of significance values suggests that transition range steepening is possibly caused by some continously-variable, non-universal process. We separate the distribution into populations corresponding to the bottom and top half percentiles, approximately distinguishing intervals best fit by 3PL from intervals for which the 2PL fit is sufficient. The notation $\chi^2_{3PL}$ and $\chi^2_{2PL}$ refers respectively to the two populations.

Figure \ref{fig:2}(b) shows the histogram of measured spectral indices for 3PL fits for $\alpha_I$, $\alpha_T$, and $\alpha_K$. Distributions are shown separately for ${\chi^2_{2PL}}$ and ${\chi^2_{3PL}}$. For the ${\chi^2_{3PL}}$ population, the spectral index for the transition range has a mean of $\langle\alpha^{\chi^2_{3PL}}_T\rangle=-3.9$, standard deviation of $0.42$, and range of  $[-5.8,-3.1]$. Table \ref{tab:table1} shows mean spectral indices for both the 2PL and 3PL fits to each range. The average 3PL transition range fit for the ${\chi^2_{2PL}}$ population values, $\langle\alpha^{\chi^2_{2PL}}_T\rangle=-3.18$, is significantly close to the mean kinetic range index for the 2PL fit, -2.9. 

\begin{table}
\caption{Measured mean spectral indices from 3PL and 2PL fits to both ${\chi^2_{2PL}}$ and ${\chi^2_{3PL}}$ populations.\label{tab:table1}}
\begin{ruledtabular}
\begin{tabular}{c|cc|ccc}
 &\multicolumn{2}{c|}{2PL Fit}&\multicolumn{3}{c}{3PL Fit}\\
 Pop.&$\langle\alpha_I\rangle$&$\langle\alpha_K\rangle$&$\langle\alpha_I\rangle$&$\langle\alpha_T\rangle$&$\langle\alpha_K\rangle$\\ \hline
 ${\chi^2_{2PL}}$&-1.7& -2.9 &-1.7& -3.1 &-2.7 \\
${\chi^2_{3PL}}$& -1.6 &-3.1& -1.6&-3.9 &-2.6\\
\end{tabular}
\end{ruledtabular}
\end{table}


 The spacecraft frequencies $f_{\rho_i}$, $f_{d_i}$ and $f_{\rho_{disp}}$ associated with wave-numbers corresponding to the ion gyro-scale $\rho_i$, inertial scale $d_i=\rho_i/\sqrt\beta_i$ and the ``dispersion scale" $\rho_{disp}=\rho_i\sqrt{(1 +T_{0e}/T_{0i})/2}$ at which the KAW become dispersive \citep{Kletzing2003}, respectively, are computed using the Taylor hypothesis $2\pi f=v_{sw} k$ where $\beta_i=v_{thi}^2/V_A^2$ and the Alfv\'en speed is $v_A=B_0/\sqrt{n_{0i}m_i\mu_0}$.

Figures \ref{fig:2}(c,d) show the probability and cumulative distributions of the break scales of the ${\chi^2_{3PL}}$ population. The measured breakpoints are normalized to the frequencies $f_{\rho_i}$, $f_{d_i}$ and $f_{\rho_{disp}}$. Figure \ref{fig:2}(c) shows that $f^{*}_{IT}$, the break from the inertial range to the anomalously steep transition range, occurs at significantly lower frequencies than any considered physical scale. Figure \ref{fig:2}(d) shows that the break between the transition to kinetic ranges, $f^{*}_{TK}$, is most similar to $f_{\rho_i}$. 

\paragraph{Physical interpretation:
dissipation.---}

\begin{figure*}
   \centering
    \includegraphics[width=6.75in]{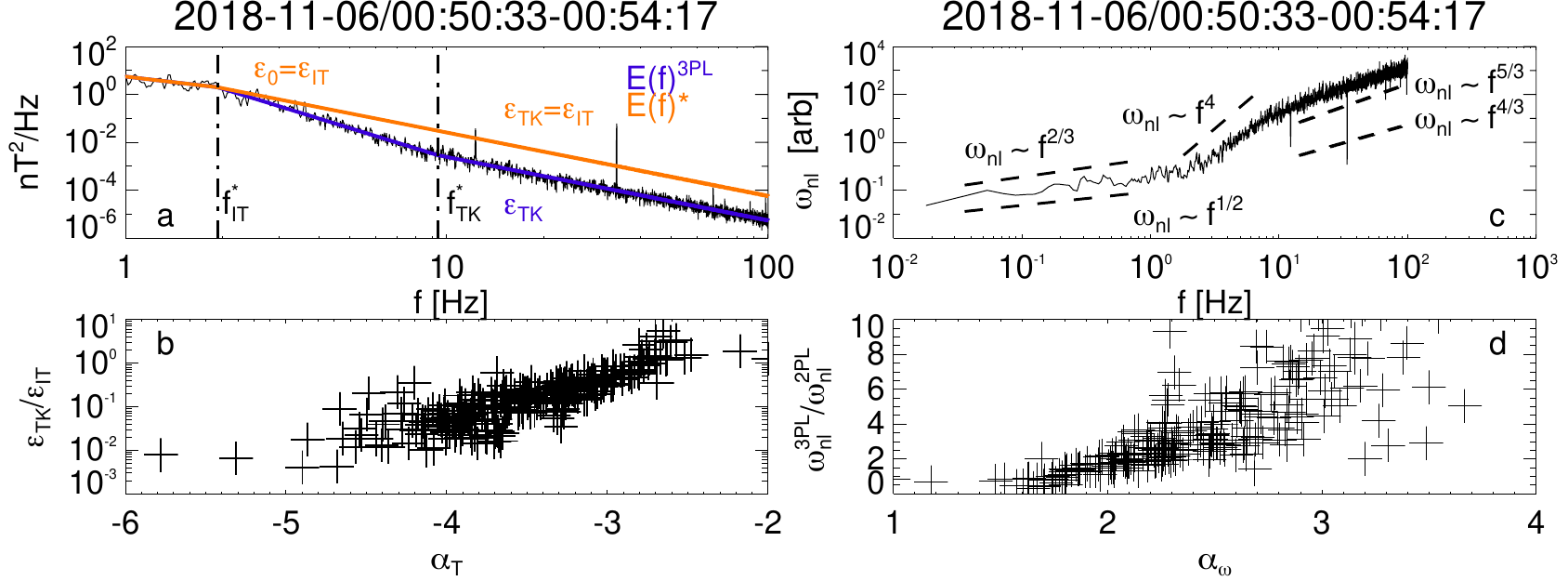}
    \caption{(a) Measured power spectrum with transition range steepening from Figure \ref{fig:1}(a) (black), with the corresponding fitted $E^{3PL}$ (blue) and synthetic $E^*$ (orange) spectra; fitted inertial-transition and transition-kinetic break scales are plotted (vertical dash-dotted lines). (b) Measured ratios of energy flux at transition-kinetic break scale relative to inertial-kinetic scale, $\epsilon_{TK}/\epsilon_{IT}$, plotted against transition range spectral index $\alpha_T$ for all intervals. (c) The dependence of the nonlinear frequency $\onl$ on spacecraft-frame frequency for interval in (a), assuming constant $\epskp$ at all frequencies. Various power-law scalings of $\onl\sim f^{\alpha_{\onl}}$ are shown with dashed lines. (d) Measured increase of $\onl$ over the transition range relative to the synthetic spectrum $E^*$, as a function of the transition range scaling index of the nonlinear frequency, $\alpha_\omega$ for all intervals.}
    \label{fig:45}
\end{figure*}
Eq.~\ref{eq:Eeps} implies that the observed steep spectra are associated with either significant dissipation or nonlinear speedup of the cascade in the transition range. First assuming that the steep spectrum is due to dissipation \citep{Denskat1983,Goldstein1994,QuataertGruzinov1999,Li2001,Howes2011,PassotSulem2015,Schreiner2017}, i.e. $\epskp$ decreases with $\kp$ across the transition range. It is necessary to then make an assumption about the baseline variation of $\tok$ with $k$. To this end, we construct from our 3PL fits a synthetic spectrum:
\begin{equation}
  {E^{*}}(f) =
    \begin{cases}
      c_I f^{\alpha_I}& \text{if } f <f_{IT}^{*} \\
           c_T f^{\alpha_K}& \text{if } f >f_{IT}^{*}. \\ 
    \end{cases}   
\end{equation}
This joins the fitted inertial-range spectrum to a synthetic spectrum with the fitted kinetic-range exponent $\alpha_k$ at the inertial-transition break $f^*_{IT}$ \footnote{Another possibility is to extend the inertial-range spectrum up to the transition-kinetic break: as can be seen, this will always result in larger fractional heating rates. We have also neglected the possibility of heating associated with the kinetic range spectral index of around $-8/3$: again, this would increase the overall heating rate. Thus, our procedure gives a lower bound on the heating.}. An example of this synthetic spectrum is shown in Figure \ref{fig:45}(a). We use $E^*$ to determine $\tok^*$ (cf. Eqs.~\ref{eq:epsE}-\ref{eq:onl}), assuming that this synthetic spectrum is what would result if $\epskp^*$ were constant. Using this synthetic $\tok^*$ and the fitted 3PL spectrum $E^{3PL}$ in Eq.~\ref{eq:Eeps} results an estimate of the fractional heating rate in the transition range relative to the synthetic spectrum,
\beq
1-\hat{Q}^* = \frac{\epsilon_{f^*_{TK}}}{\epsilon_{f^*_{IT}}}= \left(\frac{E^{3PL}(f^*_{TK})}{E^*(f^*_{TK})}\right)^{3/2}.
\eeq


Figure \ref{fig:45}(b) shows measured ratios of $\epsilon_{TK}/\epsilon_{IT}$ as a function of transition range spectral index $\alpha_T$. 
Thus, transition range spectral indices of $\alpha_T\approx -4$ may be a signature of significant ion-scale heating, corresponding in some cases to $>90\% $ of the turbulent energy flux.

\paragraph{Physical interpretation: nonlinear effects.---} Second, it is possible that the anomalously steep transition range spectrum is due to nonlinear effects which dramatically increase $\tok$, and therefore also $\onl$ (cf. Eqs.~\ref{eq:onl},\ref{eq:Eeps}) across this range. Let us now assume that $\epskp$ is a constant, and use Eq.~\ref{eq:Eeps} to determine for each interval the scaling of $\tok$, and therefore $\onl$, that matches the measured spectrum. Figure \ref{fig:45}(c) shows this for one example interval. In the inertial range, the wave-number scalings are similar to those predicted in MHD turbulence models: between $\onl \propto \kp^{2/3}$ \citep{GS95} and $\onl\propto \kp^{1/2}$ \citep{Boldyrev2006}. In the kinetic range, the scaling is again similar to predictions of the KAW turbulence models: between $\onl\propto \kp^{4/3}$ \citep{Schekochihin2009} and $\onl\propto \kp^{5/3}$ \citep{BoldyrevPerez2012}. In contrast, within the transition range, $\onl$ has a very steep scaling, which therefore may be the signature of some nonlinear process speeding up the cascade. Determining the exact mechanism behind this is beyond the scope of this Letter, though several possibilities include tearing mode physics \citep{Mallet2017b,Loureiro2017,Vech2018}, interactions between co-propagating dispersive fluctuations \citep{Voitenko2016}, or intermittent coherent structures \citep{Alexandrova2008b,Lion2016,Perrone2016}.

The fitted $E^{3PL}$ and  synthetic $E^*$ spectra allow an estimate of the increase in non-linear interactions due to transition range steepening, using Eq.~\ref{eq:epsE} and taking $\epskp$ constant. Figure \ref{fig:45}(d) shows the ratio of $\omega_{nl}^{3PL}/\omega_{nl}^{*}$ evaluated at $f^*_{TK}$. Thus, for nonlinear effects to explain the steeper spectra without dissipation, $\omega_{nl}$ must anomalously increase by a large factor of $\gtrsim 5$.


\paragraph{Discussion.---} We have performed a detailed study of the scaling properties of the turbulent fluctuation spectrum in the inner heliosphere using data from the first PSP encounter. We find that the spectrum is well-modelled by either two or three separate power-law scaling ranges. In common with previous measurements at 1AU, we find that at low frequencies, there is an ``inertial range" with a spectral index of around $-5/3$ to $-3/2$ \citep{Matthaeus1982a,Podesta2007,Chen2019}, while at high frequencies, there is a ``kinetic range" with a spectral index of around $-2.7$ \citep{Alexandrova2008,Sahraoui2009,Chen2010a}. Between these, there often appears an anomalously steep ``transition range", with a highly variable spectrum with mean spectral index around $-4$. When observed, this transition range begins at scales much larger than the characteristic ion-kinetic scales, but ends at scales comparable to the ion gyroradius. Our preliminary analysis has not identified clear correlations between signatures of the transition range and background plasma properties critical in identifying the specific process responsible: e.g. $\delta B/B_0$, $\beta_{i,e}$, $T_{0e}/T_{0i}$.

We show that this steep transition range corresponds to either significant dissipation of the turbulence into heat, or a dramatic nonlinear speedup of the cascade. We use a synthetic spectrum which has the transition range removed to show that, if dissipation is the dominant effect, the mean transition-range spectral index of $-3.9$ corresponds to 90\% of the turbulent energy flux being dissipated into heat over this range, indicating significant ion-scale heating. One important candidate mechanism which could cause this dissipation is stochastic heating \citep{Chandran2010}. Alternatively, if nonlinear speedup of the cascade is the dominant effect, we show that this means that the nonlinear frequency (inverse cascade time) of the turbulence must increase by a large factor of $\gtrsim 5$ relative to the case without a transition range. Some possibilities that could cause this effect include reconnection onset and the associated loss of dynamic alignment \citep{Mallet2017b,Vech2018}, nonlinear interactions between co-propagating waves \citep{Voitenko2016}, or the presence of intermittent coherent structures \citep{Alexandrova2008b,Lion2016,Perrone2016}.

Further work is needed to distinguish between these mechanisms. The analysis of this Letter shows that, whichever of these explanations turns out to be correct, in the inner heliosphere the observed steepness of the transition range spectrum has dramatic and important effects on the dynamics of collisionless plasma turbulence.

\bibliography{prl.bib}

\providecommand{\noopsort}[1]{}\providecommand{\singleletter}[1]{#1}%
\begin{thebibliography}{57}
\providecommand{\natexlab}[1]{#1}
\providecommand{\url}[1]{\texttt{#1}}
\expandafter\ifx\csname urlstyle\endcsname\relax
  \providecommand{\doi}[1]{doi: #1}\else
  \providecommand{\doi}{doi: \begingroup \urlstyle{rm}\Url}\fi

\bibitem[{Richardson} et~al.(1995){Richardson}, {Paularena}, {Lazarus}, and
  {Belcher}]{Richardson1995}
John~D. {Richardson}, Karolen~I. {Paularena}, Alan~J. {Lazarus}, and John~W.
  {Belcher}.
\newblock {Radial evolution of the solar wind from IMP 8 to Voyager 2}.
\newblock \emph{Geophys. Research Letters}, 22\penalty0 (4):\penalty0 325--328,
  Feb 1995.
\newblock \doi{10.1029/94GL03273}.

\bibitem[{Quataert}(1998)]{Quataert1998}
Eliot {Quataert}.
\newblock {Particle Heating by Alfv{\'e}nic Turbulence in Hot Accretion Flows}.
\newblock \emph{The Astrophysical Journal}, 500\penalty0 (2):\penalty0
  978--991, Jun 1998.
\newblock \doi{10.1086/305770}.

\bibitem[{Cranmer}(2000)]{Cranmer2000}
Steven~R. {Cranmer}.
\newblock {Ion Cyclotron Wave Dissipation in the Solar Corona: The Summed
  Effect of More than 2000 Ion Species}.
\newblock \emph{The Astrophysical Journal}, 532\penalty0 (2):\penalty0
  1197--1208, Apr 2000.
\newblock \doi{10.1086/308620}.

\bibitem[{Cranmer} and {van Ballegooijen}(2003)]{CranmervanBallegooijen2003}
S.~R. {Cranmer} and A.~A. {van Ballegooijen}.
\newblock {Alfv{\'e}nic Turbulence in the Extended Solar Corona: Kinetic
  Effects and Proton Heating}.
\newblock \emph{The Astrophysical Journal}, 594\penalty0 (1):\penalty0
  573--591, Sep 2003.
\newblock \doi{10.1086/376777}.

\bibitem[{Zhuravleva} et~al.(2014){Zhuravleva}, {Churazov}, {Schekochihin},
  {Allen}, {Ar{\'e}valo}, {Fabian}, {Forman}, {Sanders}, {Simionescu},
  {Sunyaev}, {Vikhlinin}, and {Werner}]{Zhuravleva2014}
I.~{Zhuravleva}, E.~{Churazov}, A.~A. {Schekochihin}, S.~W. {Allen},
  P.~{Ar{\'e}valo}, A.~C. {Fabian}, W.~R. {Forman}, J.~S. {Sanders},
  A.~{Simionescu}, R.~{Sunyaev}, A.~{Vikhlinin}, and N.~{Werner}.
\newblock {Turbulent heating in galaxy clusters brightest in X-rays}.
\newblock \emph{\nat}, 515\penalty0 (7525):\penalty0 85--87, Nov 2014.
\newblock \doi{10.1038/nature13830}.

\bibitem[{Chen} et~al.(2019){Chen}, {Klein}, and {Howes}]{Chen2019}
C.~H.~K. {Chen}, K.~G. {Klein}, and G.~G. {Howes}.
\newblock {Evidence for electron Landau damping in space plasma turbulence}.
\newblock \emph{Nature Communications}, 10:\penalty0 740, Feb 2019.
\newblock \doi{10.1038/s41467-019-08435-3}.

\bibitem[{Parashar} et~al.(2015){Parashar}, {Salem}, {Wicks}, {Karimabadi},
  {Gary}, and {Matthaeus}]{Parashar2015}
T.~N. {Parashar}, C.~{Salem}, R.~T. {Wicks}, H.~{Karimabadi}, S.~P. {Gary}, and
  W.~H. {Matthaeus}.
\newblock {Turbulent dissipation challenge: a community-driven effort}.
\newblock \emph{Journal of Plasma Physics}, 81\penalty0 (5):\penalty0
  905810513, October 2015.
\newblock \doi{10.1017/S0022377815000860}.

\bibitem[{Chandran} et~al.(2011){Chandran}, {Dennis}, {Quataert}, and
  {Bale}]{Chandran2011}
Benjamin D.~G. {Chandran}, Timothy~J. {Dennis}, Eliot {Quataert}, and Stuart~D.
  {Bale}.
\newblock {Incorporating Kinetic Physics into a Two-fluid Solar-wind Model with
  Temperature Anisotropy and Low-frequency Alfv{\'e}n-wave Turbulence}.
\newblock \emph{The Astrophysical Journal}, 743\penalty0 (2):\penalty0 197, Dec
  2011.
\newblock \doi{10.1088/0004-637X/743/2/197}.

\bibitem[{Chen}(2016)]{Chen2016}
C.~H.~K. {Chen}.
\newblock {Recent progress in astrophysical plasma turbulence from solar wind
  observations}.
\newblock \emph{Journal of Plasma Physics}, 82\penalty0 (6):\penalty0
  535820602, Dec 2016.
\newblock \doi{10.1017/S0022377816001124}.

\bibitem[{Matthaeus} and {Goldstein}(1982)]{Matthaeus1982a}
W.~H. {Matthaeus} and M.~L. {Goldstein}.
\newblock {Measurement of the rugged invariants of magnetohydrodynamic
  turbulence in the solar wind}.
\newblock \emph{Journal of Geophysical Research}, 87:\penalty0 6011--6028, Aug
  1982.
\newblock \doi{10.1029/JA087iA08p06011}.

\bibitem[{Podesta} et~al.(2007){Podesta}, {Roberts}, and
  {Goldstein}]{Podesta2007}
J.~J. {Podesta}, D.~A. {Roberts}, and M.~L. {Goldstein}.
\newblock {Spectral Exponents of Kinetic and Magnetic Energy Spectra in Solar
  Wind Turbulence}.
\newblock \emph{ApJ}, 664:\penalty0 543--548, July 2007.
\newblock \doi{10.1086/519211}.

\bibitem[{Goldreich} and {Sridhar}(1995)]{GS95}
P.~{Goldreich} and S.~{Sridhar}.
\newblock {Toward a Theory of Interstellar Turbulence. II. Strong Alfvenic
  Turbulence}.
\newblock \emph{The Astrophysical Journal}, 438:\penalty0 763, Jan 1995.
\newblock \doi{10.1086/175121}.

\bibitem[{Boldyrev}(2006)]{Boldyrev2006}
S.~{Boldyrev}.
\newblock {Spectrum of Magnetohydrodynamic Turbulence}.
\newblock \emph{Physical Review Letters}, 96\penalty0 (11):\penalty0 115002,
  March 2006.
\newblock \doi{10.1103/PhysRevLett.96.115002}.

\bibitem[{Chandran} et~al.(2015){Chandran}, {Schekochihin}, and
  {Mallet}]{Chandran2015}
B.~D.~G. {Chandran}, A.~A. {Schekochihin}, and A.~{Mallet}.
\newblock {Intermittency and Alignment in Strong RMHD Turbulence}.
\newblock \emph{The Astrophysical Journal}, 807\penalty0 (1):\penalty0 39, Jul
  2015.
\newblock \doi{10.1088/0004-637X/807/1/39}.

\bibitem[{Mallet} and {Schekochihin}(2017)]{MalletSchekochihin2017}
A.~{Mallet} and A.~A. {Schekochihin}.
\newblock {A statistical model of three-dimensional anisotropy and
  intermittency in strong Alfv{\'e}nic turbulence}.
\newblock \emph{MNRAS}, 466\penalty0 (4):\penalty0 3918--3927, Apr 2017.
\newblock \doi{10.1093/mnras/stw3251}.

\bibitem[{Alexandrova} et~al.(2008){Alexandrova}, {Carbone}, {Veltri}, and
  {Sorriso-Valvo}]{Alexandrova2008}
O.~{Alexandrova}, V.~{Carbone}, P.~{Veltri}, and L.~{Sorriso-Valvo}.
\newblock {Small-Scale Energy Cascade of the Solar Wind Turbulence}.
\newblock \emph{The Astrophysical Journal}, 674\penalty0 (2):\penalty0
  1153--1157, Feb 2008.
\newblock \doi{10.1086/524056}.

\bibitem[{Sahraoui} et~al.(2009){Sahraoui}, {Goldstein}, {Robert}, and
  {Khotyaintsev}]{Sahraoui2009}
F.~{Sahraoui}, M.~L. {Goldstein}, P.~{Robert}, and Yu.~V. {Khotyaintsev}.
\newblock {Evidence of a Cascade and Dissipation of Solar-Wind Turbulence at
  the Electron Gyroscale}.
\newblock \emph{Physical Review Letters}, 102\penalty0 (23):\penalty0 231102,
  Jun 2009.
\newblock \doi{10.1103/PhysRevLett.102.231102}.

\bibitem[{Chen} et~al.(2010){Chen}, {Horbury}, {Schekochihin}, {Wicks},
  {Alexandrova}, and {Mitchell}]{Chen2010a}
C.~H.~K. {Chen}, T.~S. {Horbury}, A.~A. {Schekochihin}, R.~T. {Wicks},
  O.~{Alexandrova}, and J.~{Mitchell}.
\newblock {Anisotropy of Solar Wind Turbulence between Ion and Electron
  Scales}.
\newblock \emph{Physical Review Letters}, 104\penalty0 (25):\penalty0 255002,
  Jun 2010.
\newblock \doi{10.1103/PhysRevLett.104.255002}.

\bibitem[Chen et~al.(2013)Chen, Boldyrev, Xia, and Perez]{Chen2013}
CHK Chen, S~Boldyrev, Q~Xia, and JC~Perez.
\newblock Nature of subproton scale turbulence in the solar wind.
\newblock \emph{Physical review letters}, 110\penalty0 (22):\penalty0 225002,
  2013.

\bibitem[{Schekochihin} et~al.(2009){Schekochihin}, {Cowley}, {Dorland},
  {Hammett}, {Howes}, {Quataert}, and {Tatsuno}]{Schekochihin2009}
A.~A. {Schekochihin}, S.~C. {Cowley}, W.~{Dorland}, G.~W. {Hammett}, G.~G.
  {Howes}, E.~{Quataert}, and T.~{Tatsuno}.
\newblock {Astrophysical Gyrokinetics: Kinetic and Fluid Turbulent Cascades in
  Magnetized Weakly Collisional Plasmas}.
\newblock \emph{The Astrophysical Journal Supplement}, 182:\penalty0 310--377,
  May 2009.
\newblock \doi{10.1088/0067-0049/182/1/310}.

\bibitem[{Boldyrev} and {Perez}(2012)]{BoldyrevPerez2012}
Stanislav {Boldyrev} and Jean~Carlos {Perez}.
\newblock {Spectrum of Kinetic-Alfv{\'e}n Turbulence}.
\newblock \emph{ApJ}, 758:\penalty0 L44, Oct 2012.
\newblock \doi{10.1088/2041-8205/758/2/L44}.

\bibitem[{Denskat} et~al.(1983){Denskat}, {Beinroth}, and
  {Neubauer}]{Denskat1983}
K.~U. {Denskat}, H.~J. {Beinroth}, and F.~M. {Neubauer}.
\newblock {Interplanetary magnetic field power spectra with frequencies from
  $2.4\times10^{-5}$ Hz to 470 Hz from HELIOS-observations during solarminimum
  conditions.}
\newblock \emph{Journal of Geophysics Zeitschrift Geophysik}, 54\penalty0
  (1):\penalty0 60--67, Jan 1983.

\bibitem[{Leamon} et~al.(1998){Leamon}, {Smith}, {Ness}, {Matthaeus}, and
  {Wong}]{Leamon1998a}
Robert~J. {Leamon}, Charles~W. {Smith}, Norman~F. {Ness}, William~H.
  {Matthaeus}, and Hung~K. {Wong}.
\newblock {Observational constraints on the dynamics of the interplanetary
  magnetic field dissipation range}.
\newblock \emph{Journal of Geophysical Research}, 103\penalty0 (A3):\penalty0
  4775--4788, Mar 1998.
\newblock \doi{10.1029/97JA03394}.

\bibitem[{Smith} et~al.(2006){Smith}, {Hamilton}, {Vasquez}, and
  {Leamon}]{Smith2006}
Charles~W. {Smith}, Kathleen {Hamilton}, Bernard~J. {Vasquez}, and Robert~J.
  {Leamon}.
\newblock {Dependence of the Dissipation Range Spectrum of Interplanetary
  Magnetic Fluctuationson the Rate of Energy Cascade}.
\newblock \emph{The Astrophysical Journal Letters}, 645\penalty0 (1):\penalty0
  L85--L88, Jul 2006.
\newblock \doi{10.1086/506151}.

\bibitem[{Sahraoui} et~al.(2010){Sahraoui}, {Goldstein}, {Belmont}, {Canu}, and
  {Rezeau}]{Sahraoui2010}
F.~{Sahraoui}, M.~L. {Goldstein}, G.~{Belmont}, P.~{Canu}, and L.~{Rezeau}.
\newblock {Three Dimensional Anisotropic k Spectra of Turbulence at Subproton
  Scales in the Solar Wind}.
\newblock \emph{Physical Review Letters}, 105\penalty0 (13):\penalty0 131101,
  Sep 2010.
\newblock \doi{10.1103/PhysRevLett.105.131101}.

\bibitem[{Kiyani} et~al.(2015){Kiyani}, {Osman}, and {Chapman}]{Kiyani2015}
K.~H. {Kiyani}, K.~T. {Osman}, and S.~C. {Chapman}.
\newblock {Dissipation and heating in solar wind turbulence: from the macro to
  the micro and back again}.
\newblock \emph{Philosophical Transactions of the Royal Society of London
  Series A}, 373\penalty0 (2041):\penalty0 20140155--20140155, Apr 2015.
\newblock \doi{10.1098/rsta.2014.0155}.

\bibitem[{Lion} et~al.(2016){Lion}, {Alexandrova}, and {Zaslavsky}]{Lion2016}
Sonny {Lion}, Olga {Alexandrova}, and Arnaud {Zaslavsky}.
\newblock {Coherent Events and Spectral Shape at Ion Kinetic Scales in the Fast
  Solar Wind Turbulence}.
\newblock \emph{The Astrophysical Journal}, 824\penalty0 (1):\penalty0 47, Jun
  2016.
\newblock \doi{10.3847/0004-637X/824/1/47}.

\bibitem[{Chandran} et~al.(2010){Chandran}, {Li}, {Rogers}, {Quataert}, and
  {Germaschewski}]{Chandran2010}
Benjamin D.~G. {Chandran}, Bo~{Li}, Barrett~N. {Rogers}, Eliot {Quataert}, and
  Kai {Germaschewski}.
\newblock {Perpendicular Ion Heating by Low-frequency Alfv{\'e}n-wave
  Turbulence in the Solar Wind}.
\newblock \emph{The Astrophysical Journal}, 720\penalty0 (1):\penalty0
  503--515, Sep 2010.
\newblock \doi{10.1088/0004-637X/720/1/503}.

\bibitem[Mallet et~al.(2017)Mallet, Schekochihin, and Chandran]{Mallet2017b}
Alfred Mallet, Alexander~A. Schekochihin, and Benjamin D.~G. Chandran.
\newblock Disruption of alfvénic turbulence by magnetic reconnection in a
  collisionless plasma.
\newblock \emph{Journal of Plasma Physics}, 83\penalty0 (6):\penalty0
  905830609, 2017.
\newblock \doi{10.1017/S0022377817000812}.

\bibitem[{Loureiro} and {Boldyrev}(2017)]{Loureiro2017}
Nuno~F. {Loureiro} and Stanislav {Boldyrev}.
\newblock {Role of Magnetic Reconnection in Magnetohydrodynamic Turbulence}.
\newblock \emph{Physical Review Letters}, 118\penalty0 (24):\penalty0 245101,
  Jun 2017.
\newblock \doi{10.1103/PhysRevLett.118.245101}.

\bibitem[{Vech} et~al.(2018){Vech}, {Mallet}, {Klein}, and {Kasper}]{Vech2018}
Daniel {Vech}, Alfred {Mallet}, Kristopher~G. {Klein}, and Justin~C. {Kasper}.
\newblock {Magnetic Reconnection May Control the Ion-scale Spectral Break of
  Solar Wind Turbulence}.
\newblock \emph{The Astrophysical Journal Letters}, 855\penalty0 (2):\penalty0
  L27, Mar 2018.
\newblock \doi{10.3847/2041-8213/aab351}.

\bibitem[{Voitenko} and {De Keyser}(2016)]{Voitenko2016}
Yuriy {Voitenko} and Johan {De Keyser}.
\newblock {MHD-Kinetic Transition in Imbalanced Alfv{\'e}nic Turbulence}.
\newblock \emph{The Astrophysical Journal Letters}, 832\penalty0 (2):\penalty0
  L20, Dec 2016.
\newblock \doi{10.3847/2041-8205/832/2/L20}.

\bibitem[Alexandrova(2008)]{Alexandrova2008b}
O~Alexandrova.
\newblock {Solar wind vs. magnetosheath turbulence and Alfv{\'e}n vortices}.
\newblock \emph{Nonlin. Proc. Geophys.}, 15:\penalty0 95, 2008.

\bibitem[{Perrone} et~al.(2016){Perrone}, {Alexandrova}, {Mangeney},
  {Maksimovic}, {Lacombe}, {Rakoto}, {Kasper}, and {Jovanovic}]{Perrone2016}
D.~{Perrone}, O.~{Alexandrova}, A.~{Mangeney}, M.~{Maksimovic}, C.~{Lacombe},
  V.~{Rakoto}, J.~C. {Kasper}, and D.~{Jovanovic}.
\newblock {Compressive Coherent Structures at Ion Scales in the Slow Solar
  Wind}.
\newblock \emph{The Astrophysical Journal}, 826\penalty0 (2):\penalty0 196, Aug
  2016.
\newblock \doi{10.3847/0004-637X/826/2/196}.

\bibitem[{Batchelor}(1953)]{Batchelor1953}
G.~K. {Batchelor}.
\newblock \emph{{The Theory of Homogeneous Turbulence}}.
\newblock Cambridge University Press, 1953.

\bibitem[{Howes} et~al.(2008){Howes}, {Cowley}, {Dorland}, {Hammett},
  {Quataert}, and {Schekochihin}]{Howes2008}
G.~G. {Howes}, S.~C. {Cowley}, W.~{Dorland}, G.~W. {Hammett}, E.~{Quataert},
  and A.~A. {Schekochihin}.
\newblock {A model of turbulence in magnetized plasmas: Implications for the
  dissipation range in the solar wind}.
\newblock \emph{Journal of Geophysical Research (Space Physics)}, 113\penalty0
  (A5):\penalty0 A05103, May 2008.
\newblock \doi{10.1029/2007JA012665}.

\bibitem[{Howes} et~al.(2011){Howes}, {Tenbarge}, {Dorland}, {Quataert},
  {Schekochihin}, {Numata}, and {Tatsuno}]{Howes2011}
G.~G. {Howes}, J.~M. {Tenbarge}, W.~{Dorland}, E.~{Quataert}, A.~A.
  {Schekochihin}, R.~{Numata}, and T.~{Tatsuno}.
\newblock {Gyrokinetic Simulations of Solar Wind Turbulence from Ion to
  Electron Scales}.
\newblock \emph{Physical Review Letters}, 107\penalty0 (3):\penalty0 035004,
  Jul 2011.
\newblock \doi{10.1103/PhysRevLett.107.035004}.

\bibitem[{Bale} et~al.(2016){Bale}, {Goetz}, {Harvey}, {Turin}, {Bonnell},
  {Dudok de Wit}, {Ergun}, {MacDowall}, {Pulupa}, {Andre}, {Bolton},
  {Bougeret}, {Bowen}, {Burgess}, {Cattell}, {Chandran}, {Chaston}, {Chen},
  {Choi}, {Connerney}, {Cranmer}, {Diaz-Aguado}, {Donakowski}, {Drake},
  {Farrell}, {Fergeau}, {Fermin}, {Fischer}, {Fox}, {Glaser}, {Goldstein},
  {Gordon}, {Hanson}, {Harris}, {Hayes}, {Hinze}, {Hollweg}, {Horbury},
  {Howard}, {Hoxie}, {Jannet}, {Karlsson}, {Kasper}, {Kellogg}, {Kien},
  {Klimchuk}, {Krasnoselskikh}, {Krucker}, {Lynch}, {Maksimovic}, {Malaspina},
  {Marker}, {Martin}, {Martinez-Oliveros}, {McCauley}, {McComas}, {McDonald},
  {Meyer-Vernet}, {Moncuquet}, {Monson}, {Mozer}, {Murphy}, {Odom},
  {Oliverson}, {Olson}, {Parker}, {Pankow}, {Phan}, {Quataert}, {Quinn},
  {Ruplin}, {Salem}, {Seitz}, {Sheppard}, {Siy}, {Stevens}, {Summers}, {Szabo},
  {Timofeeva}, {Vaivads}, {Velli}, {Yehle}, {Werthimer}, and
  {Wygant}]{Bale2016}
S.~D. {Bale}, K.~{Goetz}, P.~R. {Harvey}, P.~{Turin}, J.~W. {Bonnell},
  T.~{Dudok de Wit}, R.~E. {Ergun}, R.~J. {MacDowall}, M.~{Pulupa}, M.~{Andre},
  M.~{Bolton}, J.-L. {Bougeret}, T.~A. {Bowen}, D.~{Burgess}, C.~A. {Cattell},
  B.~D.~G. {Chandran}, C.~C. {Chaston}, C.~H.~K. {Chen}, M.~K. {Choi}, J.~E.
  {Connerney}, S.~{Cranmer}, M.~{Diaz-Aguado}, W.~{Donakowski}, J.~F. {Drake},
  W.~M. {Farrell}, P.~{Fergeau}, J.~{Fermin}, J.~{Fischer}, N.~{Fox},
  D.~{Glaser}, M.~{Goldstein}, D.~{Gordon}, E.~{Hanson}, S.~E. {Harris}, L.~M.
  {Hayes}, J.~J. {Hinze}, J.~V. {Hollweg}, T.~S. {Horbury}, R.~A. {Howard},
  V.~{Hoxie}, G.~{Jannet}, M.~{Karlsson}, J.~C. {Kasper}, P.~J. {Kellogg},
  M.~{Kien}, J.~A. {Klimchuk}, V.~V. {Krasnoselskikh}, S.~{Krucker}, J.~J.
  {Lynch}, M.~{Maksimovic}, D.~M. {Malaspina}, S.~{Marker}, P.~{Martin},
  J.~{Martinez-Oliveros}, J.~{McCauley}, D.~J. {McComas}, T.~{McDonald},
  N.~{Meyer-Vernet}, M.~{Moncuquet}, S.~J. {Monson}, F.~S. {Mozer}, S.~D.
  {Murphy}, J.~{Odom}, R.~{Oliverson}, J.~{Olson}, E.~N. {Parker}, D.~{Pankow},
  T.~{Phan}, E.~{Quataert}, T.~{Quinn}, S.~W. {Ruplin}, C.~{Salem}, D.~{Seitz},
  D.~A. {Sheppard}, A.~{Siy}, K.~{Stevens}, D.~{Summers}, A.~{Szabo},
  M.~{Timofeeva}, A.~{Vaivads}, M.~{Velli}, A.~{Yehle}, D.~{Werthimer}, and
  J.~R. {Wygant}.
\newblock {The FIELDS Instrument Suite for Solar Probe Plus. Measuring the
  Coronal Plasma and Magnetic Field, Plasma Waves and Turbulence, and Radio
  Signatures of Solar Transients}.
\newblock \emph{Space Science Rev.}, 204:\penalty0 49--82, December 2016.
\newblock \doi{10.1007/s11214-016-0244-5}.

\bibitem[Kasper et~al.(2016)Kasper, Abiad, Austin, Balat-Pichelin, Bale,
  Belcher, Berg, Bergner, Berthomier, Bookbinder, Brodu, Caldwell, Case,
  Chandran, Cheimets, Cirtain, Cranmer, Curtis, Daigneau, Dalton, Dasgupta,
  DeTomaso, Diaz-Aguado, Djordjevic, Donaskowski, Effinger, Florinski, Fox,
  Freeman, Gallagher, Gary, Gauron, Gates, Goldstein, Golub, Gordon, Gurnee,
  Guth, Halekas, Hatch, Heerikuisen, Ho, Hu, Johnson, Jordan, Korreck, Larson,
  Lazarus, Li, Livi, Ludlam, Maksimovic, McFadden, Marchant, Maruca, McComas,
  Messina, Mercer, Park, Peddie, Pogorelov, Reinhart, Richardson, Robinson,
  Rosen, Skoug, Slagle, Steinberg, Stevens, Szabo, Taylor, Tiu, Turin, Velli,
  Webb, Whittlesey, Wright, Wu, and Zank]{Kasper2016}
Justin~C. Kasper, Robert Abiad, Gerry Austin, Marianne Balat-Pichelin,
  Stuart~D. Bale, John~W. Belcher, Peter Berg, Henry Bergner, Matthieu
  Berthomier, Jay Bookbinder, Etienne Brodu, David Caldwell, Anthony~W. Case,
  Benjamin D.~G. Chandran, Peter Cheimets, Jonathan~W. Cirtain, Steven~R.
  Cranmer, David~W. Curtis, Peter Daigneau, Greg Dalton, Brahmananda Dasgupta,
  David DeTomaso, Millan Diaz-Aguado, Blagoje Djordjevic, Bill Donaskowski,
  Michael Effinger, Vladimir Florinski, Nichola Fox, Mark Freeman, Dennis
  Gallagher, S.~Peter Gary, Tom Gauron, Richard Gates, Melvin Goldstein, Leon
  Golub, Dorothy~A. Gordon, Reid Gurnee, Giora Guth, Jasper Halekas, Ken Hatch,
  Jacob Heerikuisen, George Ho, Qiang Hu, Greg Johnson, Steven~P. Jordan,
  Kelly~E. Korreck, Davin Larson, Alan~J. Lazarus, Gang Li, Roberto Livi,
  Michael Ludlam, Milan Maksimovic, James~P. McFadden, William Marchant,
  Bennet~A. Maruca, David~J. McComas, Luciana Messina, Tony Mercer, Sang Park,
  Andrew~M. Peddie, Nikolai Pogorelov, Matthew~J. Reinhart, John~D. Richardson,
  Miles Robinson, Irene Rosen, Ruth~M. Skoug, Amanda Slagle, John~T. Steinberg,
  Michael~L. Stevens, Adam Szabo, Ellen~R. Taylor, Chris Tiu, Paul Turin, Marco
  Velli, Gary Webb, Phyllis Whittlesey, Ken Wright, S.~T. Wu, and Gary Zank.
\newblock Solar wind electrons alphas and protons (sweap) investigation: Design
  of the solar wind and coronal plasma instrument suite for solar probe plus.
\newblock \emph{Space Science Reviews}, 204\penalty0 (1):\penalty0 131--186,
  Dec 2016.
\newblock ISSN 1572-9672.
\newblock \doi{10.1007/s11214-015-0206-3}.
\newblock URL \url{https://doi.org/10.1007/s11214-015-0206-3}.

\bibitem[Fox et~al.(2016)Fox, Velli, Bale, Decker, Driesman, Howard, Kasper,
  Kinnison, Kusterer, Lario, Lockwood, McComas, Raouafi, and Szabo]{Fox2016}
N.~J. Fox, M.~C. Velli, S.~D. Bale, R.~Decker, A.~Driesman, R.~A. Howard, J.~C.
  Kasper, J.~Kinnison, M.~Kusterer, D.~Lario, M.~K. Lockwood, D.~J. McComas,
  N.~E. Raouafi, and A.~Szabo.
\newblock The solar probe plus mission: Humanity's first visit to our star.
\newblock \emph{Space Science Reviews}, 204\penalty0 (1):\penalty0 7--48, Dec
  2016.
\newblock ISSN 1572-9672.
\newblock \doi{10.1007/s11214-015-0211-6}.
\newblock URL \url{https://doi.org/10.1007/s11214-015-0211-6}.

\bibitem[{Martinovi{\'c}} et~al.(2019){Martinovi{\'c}}, {Klein}, {Kasper},
  {Case}, {Korreck}, {Larson}, {Livi}, {Stevens}, {Whittlesey}, {Chandran},
  {Alterman}, {Huang}, {Chen}, {Bale}, {Pulupa}, {Malaspina}, {Bonnell},
  {Harvey}, {Goetz}, {Dudok de Wit}, and {MacDowall}]{Martinovic2019}
Mihailo~M. {Martinovi{\'c}}, Kristopher~G. {Klein}, Justin~C. {Kasper},
  Anthony~W. {Case}, Kelly~E. {Korreck}, Davin {Larson}, Roberto {Livi},
  Michael {Stevens}, Phyllis {Whittlesey}, Benjamin D.~G. {Chandran}, Ben~L.
  {Alterman}, Jia {Huang}, Christopher H.~K. {Chen}, Stuart~D. {Bale}, Marc
  {Pulupa}, David~M. {Malaspina}, John~W. {Bonnell}, Peter~R. {Harvey}, Keith
  {Goetz}, Thierry {Dudok de Wit}, and Robert~J. {MacDowall}.
\newblock {The Enhancement of Proton Stochastic Heating in the near-Sun Solar
  Wind}.
\newblock \emph{arXiv e-prints}, art. arXiv:1912.02653, Dec 2019.

\bibitem[{McManus} et~al.(2019){McManus}, {Bowen}, {Mallet}, {Chen},
  {Chandran}, {Bale}, {Larson}, {Dudok de Wit}, {Kasper}, {Stevens},
  {Whittlesey}, {Livi}, {Korreck}, {Goetz}, {Harvey}, {Pulupa}, {MacDowall},
  {Malaspina}, {Case}, and {Bonnell}]{McManus2019}
Michael~D. {McManus}, Trevor~A. {Bowen}, Alfred {Mallet}, Christopher H.~K.
  {Chen}, Benjamin D.~G. {Chandran}, Stuart~D. {Bale}, Davin~E. {Larson},
  Thierry {Dudok de Wit}, Justin~C. {Kasper}, Michael {Stevens}, Phyllis
  {Whittlesey}, Roberto {Livi}, Kelly~E. {Korreck}, Keith {Goetz}, Peter~R.
  {Harvey}, Marc {Pulupa}, Robert~J. {MacDowall}, David~M. {Malaspina},
  Anthony~W. {Case}, and John~W. {Bonnell}.
\newblock {Cross Helicity Reversals In Magnetic Switchbacks}.
\newblock \emph{arXiv e-prints}, art. arXiv:1912.07823, Dec 2019.

\bibitem[Bale et~al.(2019)Bale, Badman, Bonnell, Bowen, Burgess, Case, Cattell,
  Chandran, Chaston, Chen, et~al.]{Bale2019}
SD~Bale, ST~Badman, JW~Bonnell, TA~Bowen, D~Burgess, AW~Case, CA~Cattell, BDG
  Chandran, CC~Chaston, CHK Chen, et~al.
\newblock Highly structured slow solar wind emerging from an equatorial coronal
  hole.
\newblock \emph{Nature}, pages 1--6, 2019.

\bibitem[{Dudok de Wit} et~al.(2019){Dudok de Wit}, {Krasnoselskikh}, {Bale},
  {Bonnell}, {Bowen}, {Chen}, {Froment}, {Goetz}, {Harvey}, {Krishna
  Jagarlamudi}, {Larosa}, {MacDowall}, {Malaspina}, {Matthaeus}, {Pulupa},
  {Velli}, and {Whittlesey}]{Duan2019}
Thierry {Dudok de Wit}, Vladimir~V. {Krasnoselskikh}, Stuart~D. {Bale}, John~W.
  {Bonnell}, Trevor~A. {Bowen}, Christopher H.~K. {Chen}, Clara {Froment},
  Keith {Goetz}, Peter~R. {Harvey}, Vamsee {Krishna Jagarlamudi}, Andrea
  {Larosa}, Robert~J. {MacDowall}, David~M. {Malaspina}, William~H.
  {Matthaeus}, Marc {Pulupa}, Marco {Velli}, and Phyllis~L. {Whittlesey}.
\newblock {Switchbacks in the near-Sun magnetic field: long memory and impact
  on the turbulence cascade}.
\newblock \emph{arXiv e-prints}, art. arXiv:1912.02856, Dec 2019.

\bibitem[{Vech} et~al.(2019){Vech}, {Kasper}, {Klein}, {Huang}, {Stevens},
  {Chen}, {Case}, {Korreck}, {Bale}, {Bowen}, {Whittlesey}, {Livi}, {Larson},
  {Malaspina}, {Pulupa}, {Bonnell}, {Harvey}, {Goetz}, {Dudok de Wit}, and
  {MacDowall}]{Vech2019}
Daniel {Vech}, Justin~C. {Kasper}, Kristopher~G. {Klein}, Jia {Huang},
  Michael~L. {Stevens}, Christopher H.~K. {Chen}, Anthony~W. {Case}, Kelly
  {Korreck}, Stuart~D. {Bale}, Trevor~A. {Bowen}, Phyllis~L. {Whittlesey},
  Roberto {Livi}, Davin~E. {Larson}, David {Malaspina}, Marc {Pulupa}, John
  {Bonnell}, Peter {Harvey}, Keith {Goetz}, Thierry {Dudok de Wit}, and Robert
  {MacDowall}.
\newblock {Kinetic Scale Spectral Features of Cross Helicity and Residual
  Energy in the Inner Heliosphere}.
\newblock \emph{arXiv e-prints}, art. arXiv:1912.07719, Dec 2019.

\bibitem[{Bowen} et~al.(2020){Bowen}, {Bale}, {Bonnell}, {Goetz}, {Goodrich},
  {Gruesbeck}~J.and~{Harvey}, {Jannet}, {Koval}, {MacDowall}, {Malaspina},
  {Pulupa}, {Revillet}, {Sheppard}, and {Szabo}]{Bowen2020}
T.A. {Bowen}, S.D. {Bale}, T.~{Bonnell}, J.W.and~{Dudok de Wit}, K.~{Goetz},
  K.~{Goodrich}, P.R. {Gruesbeck}~J.and~{Harvey}, G.~{Jannet}, A.~{Koval}, R.J.
  {MacDowall}, D.M. {Malaspina}, M.~{Pulupa}, C.~{Revillet}, D.~{Sheppard}, and
  A.~{Szabo}.
\newblock {A Merged Search-Coil and Fluxgate Magnetometer Data Product for
  Parker Solar Probe FIELDS}.
\newblock \emph{arXiv e-prints}, art. arXiv, Jan 2020.

\bibitem[{Case} and {et al.}(2020)]{Case2020}
Anthony {Case} and {et al.}
\newblock \emph{ApJS}, 2020.

\bibitem[{Halekas} and {et al.}(2020)]{Halekas2020}
Jasper {Halekas} and {et al.}
\newblock \emph{ApJS}, 2020.

\bibitem[{Bowen} et~al.(2019){Bowen}, {Mallet}, {Huang}, {Klein}, {Malaspina},
  {Stevens}, {Bale}, {Bonnell}, {Case}, {Chandran}, {Chaston}, {Chen}, {Dudok
  de Wit}, {Goetz}, {Harvey}, {Howes}, {Kasper}, {Korreck}, {Larson}, {Livi},
  {MacDowall}, {McManus}, {Pulupa}, {Verniero}, and {Whittlesey}]{Bowen2019a}
Trevor {Bowen}, Alfred {Mallet}, Jia {Huang}, Kristopher~G. {Klein}, David~M.
  {Malaspina}, Michael~L. {Stevens}, Stuart~D. {Bale}, John~W. {Bonnell},
  Anthony~W. {Case}, Benjamin~D. {Chandran}, Christopher {Chaston},
  Christopher~H. {Chen}, Thierry {Dudok de Wit}, Keith {Goetz}, Peter~R.
  {Harvey}, Gregory~G. {Howes}, Justin~C. {Kasper}, Kelly {Korreck}, Davin~E.
  {Larson}, Roberto {Livi}, Robert~J. {MacDowall}, Michael {McManus}, Marc
  {Pulupa}, J~{Verniero}, and Phyllis {Whittlesey}.
\newblock {Ion Scale Electromagnetic Waves in the Inner Heliosphere}.
\newblock \emph{arXiv e-prints}, art. arXiv:1912.02361, Dec 2019.

\bibitem[{Howes} et~al.(2006){Howes}, {Cowley}, {Dorland}, {Hammett},
  {Quataert}, and {Schekochihin}]{Howes2006}
Gregory~G. {Howes}, Steven~C. {Cowley}, William {Dorland}, Gregory~W.
  {Hammett}, Eliot {Quataert}, and Alexand er~A. {Schekochihin}.
\newblock {Astrophysical Gyrokinetics: Basic Equations and Linear Theory}.
\newblock \emph{The Astrophysical Journal}, 651\penalty0 (1):\penalty0
  590--614, Nov 2006.
\newblock \doi{10.1086/506172}.

\bibitem[Press et~al.(1992)Press, Teukolsky, Vetterling, and Flannery]{Press}
William~H. Press, Saul~A. Teukolsky, William~T. Vetterling, and Brian~P.
  Flannery.
\newblock \emph{Numerical Recipes in C (2Nd Ed.): The Art of Scientific
  Computing}.
\newblock Cambridge University Press, New York, NY, USA, 1992.
\newblock ISBN 0-521-43108-5.

\bibitem[{Kletzing} et~al.(2003){Kletzing}, {Bounds}, {Martin-Hiner},
  {Gekelman}, and {Mitchell}]{Kletzing2003}
C.~A. {Kletzing}, S.~R. {Bounds}, J.~{Martin-Hiner}, W.~{Gekelman}, and
  C.~{Mitchell}.
\newblock {Measurements of the Shear Alfv{\'e}n Wave Dispersion for Finite
  Perpendicular Wave Number}.
\newblock \emph{Physical Review Letters}, 90\penalty0 (3):\penalty0 035004, Jan
  2003.
\newblock \doi{10.1103/PhysRevLett.90.035004}.

\bibitem[{Goldstein} et~al.(1994){Goldstein}, {Roberts}, and
  {Fitch}]{Goldstein1994}
M.~L. {Goldstein}, D.~A. {Roberts}, and C.~A. {Fitch}.
\newblock {Properties of the fluctuating magnetic helicity in the inertial and
  dissipation ranges of solar wind turbulence}.
\newblock \emph{Journal of Geophysical Research}, 99\penalty0 (A6):\penalty0
  11519--11538, Jun 1994.
\newblock \doi{10.1029/94JA00789}.

\bibitem[{Quataert} and {Gruzinov}(1999)]{QuataertGruzinov1999}
Eliot {Quataert} and Andrei {Gruzinov}.
\newblock {Turbulence and Particle Heating in Advection-dominated Accretion
  Flows}.
\newblock \emph{The Astrophysical Journal}, 520\penalty0 (1):\penalty0
  248--255, Jul 1999.
\newblock \doi{10.1086/307423}.

\bibitem[{Li} et~al.(2001){Li}, {Gary}, and {Stawicki}]{Li2001}
Hui {Li}, S.~Peter {Gary}, and Olaf {Stawicki}.
\newblock {On the dissipation of magnetic fluctuations in the solar wind}.
\newblock \emph{Geophys. Research Letters}, 28\penalty0 (7):\penalty0
  1347--1350, Apr 2001.
\newblock \doi{10.1029/2000GL012501}.

\bibitem[{Passot} and {Sulem}(2015)]{PassotSulem2015}
T.~{Passot} and P.~L. {Sulem}.
\newblock {A Model for the Non-universal Power Law of the Solar Wind
  Sub-ion-scale Magnetic Spectrum}.
\newblock \emph{The Astrophysical Journal Letters}, 812\penalty0 (2):\penalty0
  L37, Oct 2015.
\newblock \doi{10.1088/2041-8205/812/2/L37}.

\bibitem[{Schreiner} and {Saur}(2017)]{Schreiner2017}
Anne {Schreiner} and Joachim {Saur}.
\newblock {A Model for Dissipation of Solar Wind Magnetic Turbulence by Kinetic
  Alfv{\'e}n Waves at Electron Scales: Comparison with Observations}.
\newblock \emph{The Astrophysical Journal}, 835\penalty0 (2):\penalty0 133, Feb
  2017.
\newblock \doi{10.3847/1538-4357/835/2/133}.

\end{thebibliography}

\end{document}